\documentclass[twocolumn]{revtex4}
\usepackage{graphicx,amsmath,psfig}
\begin{document}
\title{Intermittent polaron dynamics: Born-Oppenheimer out of equilibrium}
\author{D. Mozyrsky, M. B. Hastings, and I. Martin}
\address{Theoretical Division, Los Alamos National Laboratory, Los Alamos, NM 87545, USA}
\date{Printed \today}
\begin{abstract}
We consider the non-equilibrium dynamics of a molecular level
interacting with local phonon modes in the case of a strong
polaronic shift which prevents a perturbative treatment of the
problem.  Instead, we find that in an adiabatic regime when the
electronic states react faster than the phonon modes it is
possible to provide a fully non-perturbative treatment of the
phonon dynamics including random noise and dissipation.  The
result shows intermittent switching between bistable states of the
oscillator with an effective random telegraph noise.
\end{abstract}
\pacs{PACS Numbers: XXXXX}
\maketitle

The adiabatic, or Born-Oppenheimer approximation is a key
approximation in dealing with a variety of strong-coupling
equilibrium problems in solid state, quantum chemistry and
numerous other fields. It is based on dividing an interacting
system into two coupled subsystems, one of which has a much lower
characteristic frequency than the other, such as splitting the
description of a molecule into nuclear and electronic degrees of
freedom. As a result one argues that the $inelastic$ processes are
suppressed by the difference in energy scales and therefore the
dynamics of the slower system is Hamiltonian, with an effective
potential given by the faster system.

In non-equilibrium situations, such as molecular transistors or
molecular junctions \cite{tour,mc1,mc2,mc3}, a straightforward
application of the Born-Oppenheimer approximation will, however,
lead to incorrect or, at least, incomplete physical picture. While
in many cases it is reasonable to assume that the mechanical
motion of the molecule between the leads is slow compared to the
dynamics of electrons hopping on and off the molecular orbitals,
the energy of the electronic $excitations$ (induced by the applied
bias voltage) may be comparable with the energy scales of
molecule's mechanical motion, As a result, inelastic processes may
occur, preventing a Hamiltonian description of molecule's
mechanical motion; instead, the effective molecular dynamics will
include noise and damping.

A simple model, which incorporates the essential physics of the
problem at hand can be described by the following Hamiltonian:
\begin{equation}
H_M = (\epsilon_0 +\lambda x)d^\dag d + {p^2\over 2m}
+{m\omega_0^2 x^2\over 2}\,.\label{1}
\end{equation}
\noindent Eq.~(\ref{1}) represents a mechanical oscillator coupled
to the extra charge $d^\dag d$ on a localized molecular orbital
$\epsilon_0$. We consider a case of sufficiently big $\lambda$, so
that the phonon deformation energy (or the polaronic shift)
$\epsilon_p = \lambda^2/(2m\omega_0^2) \geq \Gamma$, where
$\Gamma$ is the resonant level width defined below. The large
polaronic shift will prevent us from working perturbatively in
$\lambda$.  Instead, we will find below that it is possible to
work fully non-perturbatively in the classical dynamics of the
oscillator in an adiabatic limit where fluctuations about the
classic trajectory can be treated perturbatively. This calculation
may be viewed as a generalization of the Born-Oppenheimer
approximation to this case far from equilibrium.

The coupling of the molecule to the leads is described by the
Hamiltonian
\begin{equation}
H_T=\sum_{k, \alpha} \epsilon_{k\alpha}c^\dag_{k\alpha}c_{k\alpha}
+ \sum_{k,\alpha}\Delta_\alpha\left(d^\dag c_{k\alpha}+
c^\dag_{k\alpha}d\right) \,,\label{2}
\end{equation}
\noindent where operators $c_{k\alpha}$ and $d$ describe spinless fermions in
the left and right ($\alpha=L,R$) leads as in Fig. 1 and
and at the molecular orbital respectively. For simplicity
the hybridization matrix elements $\Delta_\alpha$ between the
leads and the molecule are assumed to be independent of the single
particle states in the leads (labelled by index $k$ in
Eq.~(\ref{2})). It is convenient to define partial effective
hybridization widths of the molecular orbital by $\Gamma_\alpha =
\pi\nu_\alpha\Delta_\alpha^2$, where $\nu_\alpha$ are densities of
states in the leads (the full hybridization width of the molecular
orbital is $\Gamma = \Gamma_L + \Gamma_R$). The chemical
potentials in the leads ($\mu_L$ and $\mu_R$) are assumed to be
biased by an external voltage $eV=\mu_L -\mu_R$. The leads are
assumed to be in zero temperature thermostats.

The full Hamiltonian, $H = H_M + H_T$, together with the boundary
conditions fixed by the chemical potentials of the leads, provides
a full description of the system. Here we are interested in the
dynamics of the molecule as well as in the statistical
characteristics of the current through the molecule under the
strongly non-equilibrium conditions specified below. In particular
we find that under the appropriate conditions the current through
the system exhibits random telegraph noise. We also explicitly
evaluate parameters of the noise in terms of the parameters of the
problem stated above.

Before going into the detailed analysis of the problem we note
that in the regime of strong electron-phonon coupling, e.g.,
$\epsilon_p\gg\Gamma$, the molecule is bistable, having metastable
states with with either 0 and 1 charge occupancy of the molecular
orbital. Indeed, for a negligibly small $\Gamma$, the energy
states of the molecule with an empty electronic orbital are $E_M^0
= \hbar\omega_0(n+1/2)$, $n=0,1,...$, while for the occupied
orbital the energies are $E_M^1 = \epsilon_0-\epsilon_p +
\hbar\omega_0(n+1/2)$, e.g., Eq.~(\ref{1}). These two states are
separated by an effective potential barrier of height
$\sim\epsilon_p$ (for $\epsilon_p\sim\epsilon_0$).

There are two possible mechanisms for the transition of the
molecule from one state to another, either the under-barrier
tunnelling of the molecule and or the absorption of
energy from the electronic subsystem. Transport of a single
electron through the system can be accompanied by the energy
transfer to the vibrational degree of freedom of the molecule,
i.e., the oscillator, by a value no greater than $V$. Therefore,
for sufficiently low voltage $V\ll\omega_0$ the system is in
equilibrium and the only available mechanism for its transition is
through quantum tunnelling.

A problem of this sort has been considered in the
literature~\cite{yu}. It has been shown that the partition
function of the system can be mapped onto that of interacting
Coulomb gas in 1 dimension. The properties of the latter are well
established based on the Renormalization Group (RG)
arguments~\cite{pwa}. The transition rate between the two
metastable states of the molecule is proportional to the fugacity
in the Coulomb gas model. For a symmetric case,
$\epsilon_0=\epsilon_p$, the rate is proportional to ${\rm
exp}(-\epsilon_p/\omega)$, which is essentially the quasiclassical
tunnelling exponent. An additional logarithmic renormalization of
the exponent comes from the orthogonality of electronic states
corresponding to different positioning of the
oscillator~\cite{yu}. It can be obtained from the RG
equations~\cite{pwa}. The case of finite, but small bias voltage
($V\le\omega_0$), has recently been considered in Ref.~\cite{mam}.

In this paper we are primarily interested in the opposite case,
$V\ge\omega_0$. This corresponds to a strongly out-of-equilibrium
situation as the molecule can now efficiently absorb energy from
electrons. If V is small compared to $\epsilon_p$, i.e., the
height of the potential barrier separating the two states, the
system can depart from a given state only as result of a
multi-quantum adsorption. A straightforward perturbation expansion
in powers of $\lambda$ clearly can not be applicable in this
limit, but the appropriate adiabatic expansion will be found
below.

We wish to compute a propagator for the density matrix of the
oscillator defined as
\begin{equation}
{\cal Z}_{\rm osc} = {\rm Tr}_{\rm el}\left[\rho_{\rm el} {\cal
T}_{\rm K}\,{\cal S}(-\infty,\infty) {\cal
S}(\infty,-\infty)\right]/{\rm Tr}\left[\rho_{\rm el}\right]\,
.\label{3}
\end{equation}
\noindent Here ${\cal S}(\infty,-\infty)$ and ${\cal
S}(-\infty,\infty)$ are scattering operators for the full system,
${\cal S}(\infty,-\infty) = \exp[{-i\int_{-\infty}^\infty}H dt]$,
${\cal T}_{\rm K}$ denotes the standard time ordering along the
Keldysh contour~\cite{kel}. The density matrix of the unperturbed
electrons is the direct product of the uncoupled density matrices
of electron reservoirs in the leads ($\rho_L$ and $\rho_R$) with
an empty electron state in the resonant level ($\rho_D=d
d^{\dagger}$), $\rho_{\rm el}= \rho_L\otimes\rho_R\otimes\rho_D$.
The trace in Eq.~(\ref{3}) is taken over the electronic degrees of
freedom.

The oscillator effective action can be
expressed in terms of a functional integral
\begin{equation}
{\cal Z}_{\rm osc} = \int {\cal D}x_c {\cal D}x_q e^{iS_0 +
S^\ast}\, .\label{4}
\end{equation}
\noindent where $S_0$ is a bare action of the oscillator, $S_0 =
\int dt [m{\ddot x}_c + m\omega_0^2 x_c] x_q$, and $S^\ast$
results from the trace over the electrons. One important
simplification that will enable us to work nonperturbatively in
$x_c$ is that $S^*$ vanishes identically for $x_q=0$, so that we
can perturbatively expand $S^{\ast}$ in powers of $x_q$ for a
given time-dependent $x_c(t)$. Using the cumulant expansion, the
perturbative expansion of $S^{\ast}$ to a given order in $x_q$ can
be expressed as a sum of connected diagrams, with Keldysh Green's
functions which are $2\times 2$ matrices in the space of forward
and return indices.  The vertices are $x_q(t) \times \sigma_z$,
where $\sigma_z={\rm diag}(1,-1)$ in Keldysh indices.  The Green's
functions solve the Dyson equation ${\hat G}^{-1} = {\hat
G}^{-1}_0 + \lambda x_c(t) \times {\bf 1}$.

The solution of this
Dyson equation is a difficult problem.
To make further progress, we exploit the smallness of the adiabatic
parameter $\Omega/V$, where $\Omega$ is a characteristic frequency of the
oscillator.
For $V\neq 0$ the unperturbed Green's function
for the $d$ electron is oscillatory in time with period $\sim
V^{-1}$.
Therefore the difference in the time arguments of each
$x_q(t)$ in a given term in $S^*$ should not
significantly exceed $V^{-1}$; define $t_0$ to be the
largest of these time arguments.
Let us set $x_c(t) = x_c(t_0)
+{\dot x_c}(t_0)(t-t_0)+...$.
Then, to compute a given term in $S^*$, we first compute the
Green's functions in the presence of a {\it fixed} $x_c(t)=x_c(t_0)$
and then perturbatively
expand the Green's functions in derivatives of ${\dot x_c}(t_0)$.
One finds that
a term with $n$ derivatives of $x_c$ is the suppressed by $(\Omega/V)^n$.

The $x_q$ component of ${\hat x}(t)$ essentially defines
deviations of the oscillator's density matrix from the diagonal
matrix. It can be easily checked that in an adiabatic
approximation contribution of a bubble containing $n$ $x_q$-
vertices scales as $(\lambda x_q/\Gamma)^n V/\Omega$ for $n>1$.
Therefore if we wish to integrate over $x_q$ in Eq.~(\ref{4}), we
have to perform an integral of a sort $\int dx_q \exp{[\beta x_q
+(V/\Omega)\eta(x_q)]}$.  In the spirit of the saddle point
approximation, the factor $(V/\Omega)$ in front again
suppresses higher powers of
$x_q$ by powers
of $(\Omega/V)$.

Thus, the same parameter $\Omega/V$ controls the higher
derivatives of $x_c$ as well as the higher powers of $x_q$. Later,
the validity of this expansion must be verified by a
self-consistency check.  We will find that in the region of
interest, i.e., in case of two wells separated by a high barrier
(see Fig. 1) $\Omega\sim\omega_0$ and therefore the expansion is
well justified far from equilibrium with $\Omega<<V$. Then, to
leading order in $\Omega/V$ the effective action in Eq.~(\ref{4})
involves three diagrams: the diagram with a single $x_q$-vertex,
the diagram with one $x_q$-vertex and one ${\dot
x}_c(t_0)(t-t_0)$-vertex, and the diagram with two $x_q$-vertices.
One has to evaluate the contribution of each of these three
diagrams with the renormalized adiabatic Green's function of the
$d$ electron, defined by ${\hat G}^{-1} = {\hat G}^{-1}_0 +
\lambda x_c\times {\bf 1}$. The off-diagonal components of ${\hat
G}$, i.e., $-i\langle d_f(t)d_r^\dag(t_0)\rangle$ and $-i\langle
d_r(t)d_f^\dag(t_0)\rangle$ can be explicitly expressed in terms
of their Fourier transforms, $(1/2\pi i)\int
d\omega\exp{[-i\omega(t-t_0)]}G_{ij}[\omega, x_c(t_0)], \
i,j=f,r$:
\begin{subequations}
\label{5}
\begin{eqnarray}
&&\hskip-8mm G_{fr}[\omega, x_c] =
2i{\Gamma_L\Theta(\mu_L-\omega)+\Gamma_R\Theta(\mu_R-\omega)\over
(\omega-\epsilon_0-\lambda x_c)^2 + \Gamma^2} \ ,\\\label{5a}
&&\hskip-8mm G_{rf}[\omega, x_c] =
-2i{\Gamma_L\Theta(\omega-\mu_L)+\Gamma_R\Theta(\omega
-\mu_R)\over (\omega-\epsilon_0-\lambda x_c)^2 + \Gamma^2} \
.\label{5b}
\end{eqnarray}
\end{subequations}
In terms of the adiabatic Green's function ${\hat G}$ the
semiclassical effective action for the oscillator can be written
as $S^\ast_{s-c} =$
\begin{equation}
\int dt[iF(x_c)x_q + iA(x_c){\dot x}_c x_q + (1/2)D(x_c)x_q^2]  \,
,\label{6}
\end{equation}
\noindent where
\begin{subequations}
\label{7}
\begin{eqnarray}
&&\hskip-8mm F(x_c) = {\lambda\over 2\pi i}\int d\omega
G_{fr}[\omega, x_c]\ ,\\\label{7a} &&\hskip-8mm A(x_c) =
{\lambda^2\over 2\pi}\int d\omega G_{fr}[\omega,
x_c]{\partial\over\partial\omega}G_{rf}[\omega, x_c] \
,\\\label{7b} &&\hskip-8mm D(x_c) = {\lambda^2\over 2\pi}\int
d\omega G_{fr}[\omega, x_c]G_{rf}[\omega, x_c] \ .\label{7c}
\end{eqnarray}
\end{subequations}
The first term in Eq.~(\ref{6}) is a classical force exerted by
electrons onto the oscillator. Therefore the effective potential
for the oscillator is $U(x)= m\omega_0^2x^2/2 + \int^x dx^\prime
F(x^\prime)$. In the bistable regime, $\epsilon_p =
\lambda^2/(2m\omega_0^2) \geq \Gamma,\ 0<\epsilon_0<\epsilon_p$,
and not too high voltage, $\omega_0<V<E_p$, $U(x)$ develops two
minima separated by a barrier.  The minima correspond to the
states of the molecule with occupied and empty orbital. Though the
approach developed in this paper allows to deal with a pretty
arbitrary situation as long as $\Omega \ll V$, here we will
concentrate on the study of the interesting bistable case. For
mathematical simplicity let us assume that the hybridizations
$\Gamma_L$ and $\Gamma_R$ with leads are equal. Then the two
minima are located at $x_0 = \epsilon_0/\lambda$ and at
$x_1=(\epsilon_0-2\epsilon_p)/\lambda$, with energies $E_0 =
-\epsilon_0^2/4\epsilon_p$ and $E_1 =
-(\epsilon_0-2\epsilon_p)^2/4\epsilon_p$, see Fig. 1, while
maximum is at $x=0$. In the limit $E_p\gg\Gamma$ the wells are
parabolic with frequencies of small oscillation  are equal to
$\omega_0$.
\begin{figure}[h]
{\centering{\psfig{figure=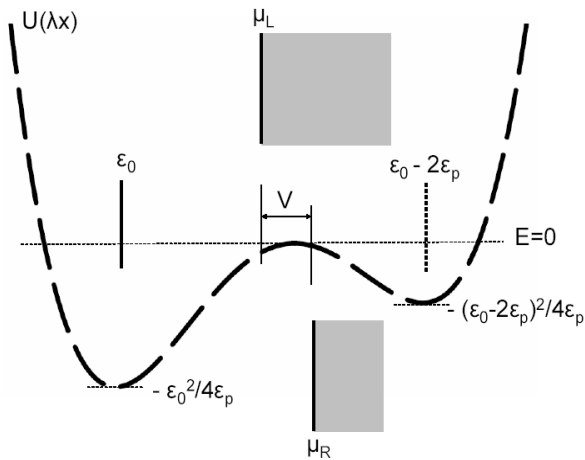,width=3.2 in,angle=90}}}
\caption{Symbolic representation of system's energetics: the
potential curve of the mechanical degree of freedom versus the
electronic energy levels.}
\end{figure}

The last two terms in $S^\ast$ in Eq.~(\ref{6}) correspond to
dissipation and fluctuation of the oscillator due to electrons. It
is convenient to decouple the last term in Eq.~(\ref{6}) by
introducing the Gaussian noise field $\xi(t)$ in the functional
integral Eq.~(\ref{4}) by means of Hubbard-Stratanovich
transformation and integrate out the $x_q$ field. As a result the
partition function ${\cal Z}_{\rm osc}$ can be explicitly related
to the distribution function for a classical particle, whose
dynamics is governed by a Langevin equation:
\begin{eqnarray}
&& \begin{array}{l} {\displaystyle{\cal Z}_{osc}=\int {\cal
D}\xi\int {\cal D}x_c \exp{\left[-\int dt {\xi^2(t)\over
2D(x_c)}\right]}\times}\\{\displaystyle{\bf \delta}_t\left[m{\ddot
x_c} - F(x_c) + A(x_c){\dot x}_c - \xi(t)\right]\ ,\phantom{\Big)}
}
\end{array} \label{71}
\end{eqnarray}
\noindent where the Gaussian noise $\xi(t)$ is white, i.e.,
$\langle\xi(t)\xi(t^\prime)\rangle = D(x)\delta(t-t^\prime)$, and
${\bf \delta}_t$ in Eq.~(\ref{71}) is a functional
$\delta$-function. From now on we will drop the unnecessary
$c$-subscript.

The damping of the oscillator defined by an effective friction
coefficient $\gamma(x) = A(x)/m$, where $A$ is given by
Eq.~(\ref{7a}), is strongest in the vicinity of the barrier,
$\gamma\,_{\rm barrier}\sim \epsilon_p (\omega_0/\Gamma)$. Away
from the barrier it decays as $\gamma(x)\sim x^{-4}$ and the
system is underdamped everywhere except for a narrow region of
width $\Gamma/\lambda$ at the barrier, where it approaches a
separatrix and thus its period diverges. Therefore its natural to
consider motion of the oscillator as a diffusion in energy space.
The rate of change of the mean energy of the oscillator over a
period depends of both dissipation and the fluctuation as $\langle
dE/dt\rangle_T = -\langle A(x){\dot x}^2\rangle_T + \langle
\xi(t){\dot x}\rangle_T$. The friction-dependent term can be
expressed in terms of the instantaneous energy of the oscillator
as $-1/T\oint dt A(x){\dot x}^2$, where integral is taken over a
period of motion $T$ of the oscillator. Since the motion of the
oscillator during one period is nearly deterministic (for small
fluctuation and dissipation), the noise introduced by the
$\langle\xi(t) {\dot x}\rangle_T$ term is also Gaussian. The first
moment of this noise is given by ${1/(2T)}\oint dt D(x)$ (the Ito
term), while the second moment is $1/T\oint dt D(x){\dot x}^2$.
The Ito term offsets the damping term so that, due to
fluctuations, the system tends to occupy states close to, but not
exactly at, the bottom of the potential curve $U$. The Langevin
equation for system's energy can be cast in a form:
\begin{eqnarray}
&& \begin{array}{l} {\displaystyle{dE\over dt} = -\gamma (E) +
\xi_E(t)}\, , \\
   {\displaystyle \langle\xi_E (t)\xi_E(t^\prime)\rangle = D(E)\delta(t-t^\prime)\,
   , \phantom{\Big)} }
\end{array} \label{8}
\end{eqnarray}
\noindent with the damping and the diffusion coefficients in the
energy space:
\begin{subequations}
\label{9}
\begin{eqnarray}
&&\hskip-8mm \gamma(E) = {2\over T}\int dx \left[A(x){\dot
x}-{D(x)\over 2\dot x}\right]\ ,\\\label{9a} &&\hskip-8mm D(E) =
{2\over T}\int dx\, D(x){\dot x} \,,\label{9b}
\end{eqnarray}
\end{subequations}
\noindent where ${\dot x} = \sqrt{2[E-U(x)]/m}$, the period of
motion $T = 2\int dx \sqrt{m/2{[E-U(x)]}}$, and the integration
runs between the turning points defined by $U(x) = E$.

Eq.~(\ref{8}) describes diffusion of the oscillator's energy in
the nonlinear potential $U$. The Fokker-Plank equation which
corresponds to dynamics in Eq.~(\ref{8}) reads:
\begin{equation}
{\partial P\over\partial t} = {\partial\over\partial
E}\left[\gamma(E)P\right]+{1\over 2}{\partial^2\over\partial E^2}
\left[D(E)P\right]\, .\label{10}
\end{equation}
\noindent In case of the two-well potential there are three
connected regions: two in each well and one above the barrier.
Solutions of the Fokker-Plank equation in each region are related
through a boundary conditions $P_0(E=0) + P_1(E=0) = P_{\rm
above}(E=0)$, and $P_0(E=0) = P_1(E=0)$, which state that when the
system reaches the separatrix, it is equally likely to be in both
wells. In equilibrium, i.e., in the absence of the net flux, the
stationary distribution for the oscillator can be found from an
equation $\gamma P_{\rm st} = -(1/2)\partial (DP_{\rm
st})/\partial E$.

The current operator can be defined as $I = ({\dot N}_L - {\dot
N}_R)/2$, where $N_\alpha = \sum_k c^\dag_{k\alpha}c_{k\alpha}, \
\alpha=L,R$ and ${\dot N}_\alpha = i[N_\alpha, H]$. In the
adiabatic limit the current expectation value can be expressed in
terms of the stationary distribution of the oscillator as $\int dE
P_{\rm st}(E)\langle I(E)\rangle$. Here $\langle I(E)\rangle =
{1/T}\oint dt \langle I(x)\rangle$, where $\langle I(x)\rangle$ is
the expectation value of the current evaluated for a fixed
position $x$ of the oscillator (brackets stand for averaging over
the electronic degrees of freedom only). Evaluation of the above
integrals is easy for $\epsilon_p \gg {\rm max}(V, \Gamma)$.  Then
the oscillator tends to be near the bottom of one of the two
potential wells with negligible probability of being near the top
of the barrier, leading to
\begin{eqnarray}
&& \begin{array}{l} {\displaystyle \langle I \rangle = I_1
n_F^\ast(\epsilon_0) + I_2 [1-n_F^\ast(\epsilon_0)]} \, , \phantom{|_{\displaystyle |_{\displaystyle |}}} \\
   {\displaystyle n_F^\ast(\epsilon_0) = {1\over 1+\exp[4(\epsilon_0 - \epsilon_p)/V]}\,
   , \phantom{\Big)} }
\end{array} \label{11}
\end{eqnarray}
\noindent where $I_1 = \Gamma^2 V/[2\pi(\epsilon_0 -
2\epsilon_p)^2]$ and $I_2 = \Gamma^2 V/[2\pi\epsilon_0^2]$ are the
currents through the molecule for the oscillator positioned at the
minima of each well. The coefficient $n_F^\ast$ has the same form
as the Fermi distribution function with temperature $V/4$ and
chemical potential $\epsilon_p$.

In this case, we find that each well has the same effective temperature
$V/4$, and thus the occupation probability of a given well in Eq.~(\ref{11})
is given by a Boltzmann weight.  In general, each well may have a different
effective temperature; in that case, the oscillator may be more likely to
be found in the higher energy state if the effective temperature of that
well is lower.

The current-current correlation function requires evaluating the
expectation value, averaged over electronic states, of $\langle
I(E,t_1) I(E,t_2)\rangle$. In the adiabatic limit the connected
part of this expectation value can be expressed in terms of the
stationary distribution of the oscillator as $\int dE P_{\rm
st}(E) \langle I^2(E) \rangle_{\rm con}$, where $I^2(E)=1/T \oint
dt \langle I^2(x) \rangle$. This corresponds to the shot noise
$\sim 2e\langle I\rangle$.

The disconnected part is
\begin{eqnarray}
\int dE_1 dE_2 P_{\rm tr}(E_1,t_1;E_2,t_2)\langle
I(E_1,t_1)\rangle\langle I(E_2,t_2)\rangle \, ,\label{13}
\end{eqnarray}
\noindent where $P_{\rm tr}(E_1,t_1;E_2,t_2)$ is the 2-point
distribution function for the oscillator. This part contains a
contribution due to the oscillations in each well, and due to the
transitions between the wells. For a sufficiently high barrier
($E_p\gg V$), the transitions (jumps) are rare events (note that
$V$ is playing role of temperature in the wells), and therefore
the oscillations and the jumps occur on different time scales. The
contribution of oscillatory motion is peaked at $\omega =
\omega_0$, with a width of the order of $\gamma(x_{\rm
min})$\cite{hmm}. The contribution of jumps into the
current-current correlator in Eq.~(\ref{13}) is $\sum_{ij}
P_{ij}(t_1 -t_2)I_i I_j P_{{\rm st}\,j}$, where $i,j=1,2$ label
the states in each well, and $P_{{\rm st}\,i}$ are the
probabilities for the oscillator to be found in each well,
$P_{{\rm st}\,i} = \int_{{\rm well}\,i} dE P_{\rm st}(E)$. The
transition probabilities $P_{ij}$ between the two wells are
exponentially decaying on time-scales $\tau_i$, where $\tau_i$ are
the mean first passage times for each well~\cite{zw}. In this
two-state approximation the correlation function corresponds to
that of a telegraph noise with spectrum $S_{II}^{\rm tel} = S^{\rm
tel}(0)/(1+\omega^2\tau^2)$, where $S^{\rm
tel}(0)=4\tau_1^{-1}\tau_2^{-1}\tau^3(I_1-I_2)^2$, and $\tau^{-1}
= \tau_1^{-1} + \tau_2^{-1}$. The mean first passage times
$\tau_i$ can be evaluated from the Fokker-Plank Eq.~(\ref{10})
based on standard techniques, i.e., by solving an equation adjoint
to Eq.~(\ref{10})~\cite{zw}. One finds:
\begin{equation}
\tau_i^{-1} =  \left({\omega_0\over V}\right)^2
I_i\exp{(-{4E_i\over V})} \, ,\label{12}
\end{equation}
\noindent where $E_1 = \epsilon_0^2/4E_p$ and $E_2 =
(\epsilon_0-2\epsilon_p)^2/4E_p$, e.g., Fig. 1. The telegraph
switching is thus highly sensitive to the bias voltage and can be
controlled by both the bias voltage and the gate voltage through
the variation of the molecular orbital energy ($\epsilon_0$)
relative to the conduction bands of electrons in the leads.

We would like to acknowledge useful discussions with M. Stepanov.
This work was supported by the US DOE. D.M. was supported, in
part, by DMR-0121146.

\end{document}